\documentclass[pdftex,twocolumn,epjc3]{svjour3} % twocolumn

\usepackage{cite}
\usepackage{lineno}
\usepackage{soul}
\usepackage{color}

\RequirePackage[T1]{fontenc}

\smartqed % flush right qed marks, e.g. at end of proof

\RequirePackage{graphicx}
\RequirePackage{mathptmx} % use Times fonts if available on your TeX system
\RequirePackage{flushend}
\RequirePackage[numbers,sort&compress]{natbib}
\RequirePackage[colorlinks,citecolor=blue,urlcolor=blue,linkcolor=blue]{hyperref}

\journalname{Eur. Phys. J. C}

\begin{document}
%\linenumbers

\title{
Virtual depth by active background suppression: 
Revisiting the cosmic muon induced background of \textsc{Gerda} Phase\,II}


\author{Christoph Wiesinger\thanksref{e1,addr1}
\and    Luciano Pandola\thanksref{addr2}
\and    Stefan Sch\"onert\thanksref{addr1}  %etc.
}

%\thankstext[$\star$]{t1}{Thanks to the title}
\thankstext{e1}{e-mail: christoph.wiesinger@tum.de}

\institute{Physik Department E15, Technische  Universit{\"a}t M{\"u}nchen, Garching, Germany\label{addr1}
     \and  INFN Laboratori Nazionali del Sud, Catania, Italy\label{addr2}
}

\date{Received: date / Accepted: date} % The correct dates will be entered by the editor

\maketitle

%----------------------------------------------------------------------------------
%------------------------------------------------------------------------- ABSTRACT
\begin{abstract}
In-situ production of radioisotopes by cosmic muon interactions may generate a non-negligible background for deep underground rare event searches.
Previous Monte Carlo studies for the \textsc{Gerda} experiment at \textsc{Lngs} identified the delayed decays of {$^{77}$Ge} and its metastable state {$^{77m}$Ge} as dominant cosmogenic background in the search for neutrinoless double beta decay of {$^{76}$Ge}.
This might limit the sensitivity of next generation experiments aiming for increased {$^{76}$Ge} mass at background-free conditions and thereby define a minimum depth requirement.
A re-evaluation of the {$^{77(m)}$Ge} background for the \textsc{Gerda} experiment has been carried out by a set of Monte Carlo simulations.
The obtained {$^{77(m)}$Ge} production rate is (0.21$\pm$0.01)~nuclei/(kg$\cdot$yr). 
After application of state-of-the-art active background suppression techniques and simple delayed coincidence cuts this corresponds to a background contribution of (2.7$\pm$0.3)$\cdot10^{-6}$ cts/(keV$\cdot$kg$\cdot$yr). %at {$Q_{\beta\beta}$} at reasonable life-time loss of <~4\%.
The suppression achieved by this strategy equals an effective muon flux reduction of more than one order of magnitude.
This virtual depth increase opens the way for next generation rare event searches.
\end{abstract}

%----------------------------------------------------------------------------------
%--------------------------------------------------------------------- INTRODUCTION
\section{Introduction}
\label{sec:intro}

In-situ production of radioactive isotopes by cosmic muon interactions constitutes a non-negligible background for rare event searches and may define a minimum depth for experiments aiming for ultra-low backgrounds \cite{MeiHime}.
In \cite{NIMA} the muon induced background for the \textsc{Gerda} (GErmanium Detector Array) experiment \cite{GerdaEPJC} at \textsc{Lngs} (Laboratori Nazionali del Gran Sasso) of \textsc{Infn} was studied.
The delayed decays of {$^{77}$Ge} and its isomeric state {$^{77m}$Ge} were identified as dominant cosmogenic background in \textsc{Gerda}'s search for neutrinoless double beta ({$0\nu\beta\beta$}) decay of {$^{76}$Ge}.
{$^{77(m)}$Ge} is formed by neutron capture on the double beta isotope {$^{76}$Ge} itself.\footnote{The notation {$^{77(m)}$Ge} represents {$^{77}$Ge} and {$^{77m}$Ge}.}
A background contribution of (1.1$\pm$0.2)$\cdot10^{-4}$ cts/(keV$\cdot$kg$\cdot$yr) at {$0\nu\beta\beta$} relevant energies before active background rejection cuts was reported.
This is well below the background of \textsc{Gerda} Phase\,II, but might constitute a significant fraction of the background budget for the next generation experiment \textsc{Legend} (Large Enriched Germanium Experiment for Neutrinoless $\beta\beta$ Decay) \cite{LegendMedex}.
\textsc{Legend} is aiming to perform a background-free search for {$0\nu\beta\beta$} decay with increased {$^{76}$Ge} mass.

In this work a re-evaluation of the {$^{77(m)}$Ge} background in \textsc{Gerda} is carried out by a set of Monte Carlo simulations. 
They use the actual Phase\,II geometry and up-to-date {$^{76}$Ge} neutron capture cross section data.
A description of the \textsc{Gerda} setup and its Monte Carlo implementation can be found in Section~\ref{sec:gerda}.
Results from {$^{77}$Ge} and {$^{77m}$Ge} decay simulations and the impact of active background suppression techniques are shown in Section~\ref{sec:ge77}.
The production of {$^{77(m)}$Ge} by cosmic muon induced neutrons is discussed in Section~\ref{sec:ncapture}.
Prospects for reduction by delayed coincidence cuts are described in Section~\ref{sec:cuts}. 
In Section~\ref{sec:others} minor cosmogenic background contributions from other sources are discussed briefly.
Final conclusions on the feasibility of a next generation {$^{76}$Ge} experiment at \textsc{Lngs} exploiting the \textsc{Gerda} passive and active shielding approach are drawn in Section~\ref{sec:concl}.

%----------------------------------------------------------------------------------
%---------------------------------------------------------------------------- GERDA
\section{\textsc{Gerda}}
\label{sec:gerda}

\textsc{Gerda} is searching for the $0\nu\beta\beta$ decay of {$^{76}$Ge}. 
High purity germanium (HPGe) detectors enriched in {$^{76}$Ge} constitute simultaneously source and detector for the search of the mono-energetic line at the Q-value of 2039~keV ({$Q_{\beta\beta}$}).
\textsc{Gerda} is situated in Hall A of the \textsc{Lngs} underground laboratory of \textsc{Infn} in Italy.
\textsc{Lngs} provides an overburden of 3500~m.w.e. and a residual muon flux of $\sim$1.25~m$^{-2}$h$^{-1}$.
The outermost part of the experiment is a 590~m$^3$ water tank.
The water efficiently acts as neutron absorber and is equipped with photomultiplier tubes (PMTs) to serve as Cherenkov veto for cosmic muons.
The muon veto system is completed by plastic scintillator panels on top of the experiment.
The water tank surrounds a cryostat filled with 64~m$^3$ of liquid argon (LAr).
The instrumentation is lowered into this cryostat from a clean room on top of the experiment via an air-tight lock system.
The first phase of \textsc{Gerda} (Phase\,I) was performed between 2011 and 2013 \cite{GerdaPRL}.
In Phase\,II 40 HPGe detectors are operated in a 7 string array configuration.
37 of them are made from isotopically enriched germanium material with an enrichment fraction of 87\% {$^{76}$Ge}.
They constitute a detector mass of 35.6~kg.
The 3 remaining detectors feature natural isotopic composition.
The LAr volume around the array is instrumented with wavelength shifting fibers coupled to silicon photomultipliers (SiPMs) \cite{Fiber1,Fiber2} and low-activity PMTs. 
The ability to detect scintillation light allows to reject backgrounds with coincident energy release in HPGe detectors and LAr \cite{LArGe}.
Results from Phase\,II were published in \cite{GerdaNature,GerdaPRL2}.
A detailed description of the \textsc{Gerda} Phase\,II setup can be found in \cite{upgrade}.

\subsection{Monte Carlo implementation}
\label{subsec:mc}

The Monte Carlo simulations for this work were carried out by the \textsc{Geant4}-based \cite{geant4_1,geant4_2,geant4_3} \textsc{MaGe} framework \cite{MaGe}.
It is jointly developed and maintained by the \textsc{Gerda} and \textsc{Majorana} \cite{Mjd} collaborations.
The \textsc{Gerda} Phase\,II implementation includes all relevant components of the experiment as well as the rock surroundings of Hall A at \textsc{Lngs}.
The simulations in this work are performed with \textsc{MaGe} built against \textsc{Geant4} 10.3.
The default physics list of \textsc{MaGe} is used.\footnote{Simulations with the reference Shielding physics list provided by \textsc{Geant4} produced equal results.}
It corresponds to the reference physics list QGSP\_BERT\_HP of \textsc{Geant4}. 
Inelastic interactions of nucleons and pions are simulated according to the theory-driven quark-gluon string (QGS) model above 20~GeV; according to the Fritiof (FTF) model between 10 and 20~GeV; and according to the Bertini cascade (BERT) model below~10 GeV.
Interactions of neutrons below 20~MeV and down to thermal energies are described accurately by the high-precision data-driven models (NeutronHP) based on the evaluated ENDF/B-VII data libraries \cite{endf}. 
The specialized low-energy models based on the Livermore data library are used for the electromagnetic interactions of electrons and $\gamma$ rays. 
The model for muon-nuclear interaction is G4MuNuclearInteraction. 
This physics list has been shown to provide good performance in handling low-energy electromagnetic interactions as well as hadronic showers initiated by high-energy cosmic muons \cite{MaGe}.  

The production of isotopes in muon induced cascades can be either due to the hadronic (or lepton-nuclear) interactions within the hadronic shower, or to secondary neutrons emerging from the shower. 
The former is the dominant mechanism for the production of nuclei (A,Z) "far away'' with respect to the target nucleus (e.g. {$^{60}$Co} production from Ge). 
The production of "nearby'' nuclei is mostly ascribed to secondary neutron interactions, as radiative capture (n,$\gamma$) and inelastic scattering (n,X), e.g. (n,p). 
This is the case of the most relevant species discussed in this paper, namely {$^{77(m)}$Ge} from {$^{76}$Ge}, {$^{41}$Ar} and {$^{40}$Cl} from {$^{40}$Ar}.

The evaluation of the neutron-mediated production yield involves two ingredients, namely the generation of neutrons from the muon induced showers and the neutron tracking down to their capture/interaction, which may happen at thermal energies. 
The capability of \textsc{Geant4} to provide a realistic estimate of the muon induced neutron production in underground sites has been widely discussed in the literature (see e.g. \cite{Araujo:2004rv,Kudryavtsev:2008fi,Reichhart:2013xkd}). 
The neutron yield predicted by \textsc{Geant4} for muon induced interactions in a low-A target, as organic liquid scintillator, was found to agree with the experimental results from KamLAND within 10\% \cite{Abe:2009aa}. 
As for neutron production in high-A materials, the recent work of \cite{Abt:2016nor,Du:2018mzh} reports that \textsc{Geant4} under-estimates it by a factor of 3-4 in Pb for cosmic muons at shallow depth ($\langle E_{\mu} \rangle$=7~GeV). 
However, \cite{Reichhart:2013xkd} shows that \textsc{Geant4} slightly over-estimates the neutron production in Pb in a deep underground laboratory ($\langle E_{\mu} \rangle$=~260 GeV) by 25\% and that the choice of the physics list has a <~5\% impact.
The weak dependency on the physics list was also observed in \cite{MaGe}: the differences in neutron yield from high-energy muons in metallic germanium were found to be within 15\%.
As the configuration in \cite{Reichhart:2013xkd} is more representative of the muon energy spectrum at \textsc{Lngs} ($\langle E_{\mu} \rangle$=270 GeV) than in \cite{Du:2018mzh}, the systematic uncertainty on the muon-induced neutron production is taken to be 25\%. 
This is also reasonably conservative, as the \textsc{Gerda} setup is composed entirely by low-A materials, with water and argon being the most relevant.

%%%%%%%%%%%%%%%%%%%%%%%%%%%%%%%%%%%%%%%%%%%%%%%%%%%%%%%%%%%%%%%%%%%% figure %%%%
\begin{figure*}[t]
\centering
\includegraphics[width=\textwidth]{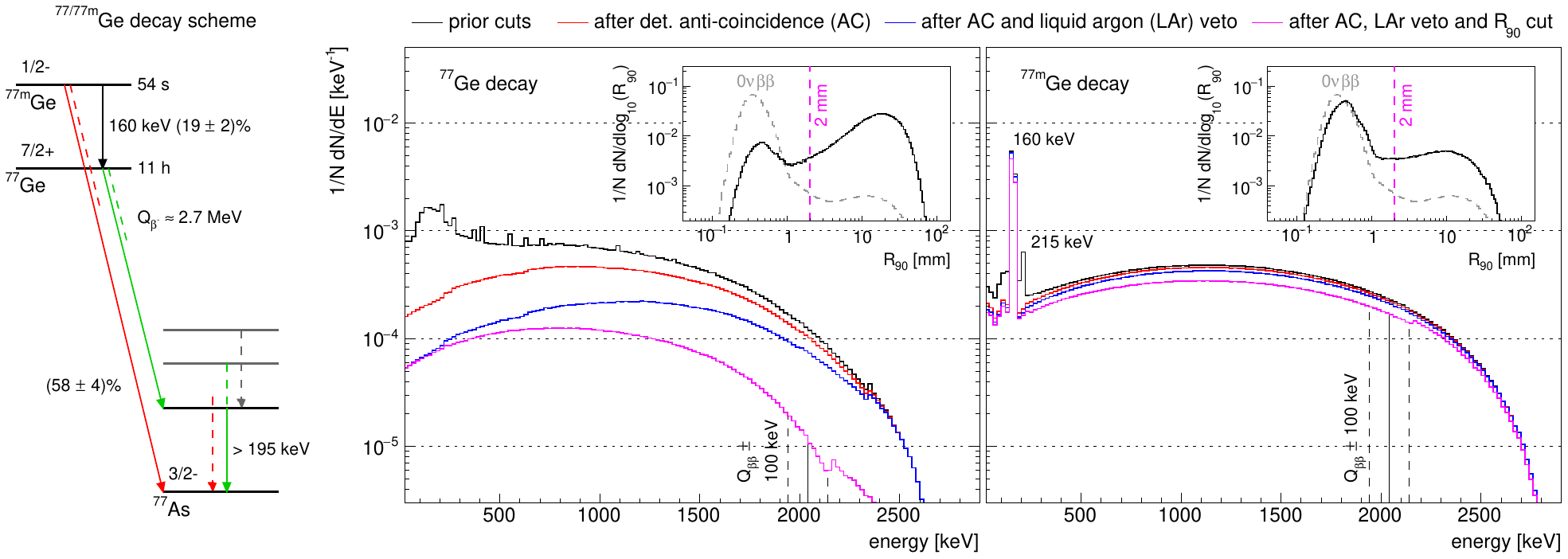}
\caption{
Left: Simplified decay scheme for {$^{77}$Ge} and {$^{77m}$Ge}. 
Both isotopes undergo $\beta$-decay to {$^{77}$As}. 
{$^{77m}$Ge} undergoes mainly pure ground-state decay without the release of coincident $\gamma$'s.
Middle/Right: Spectra for {$^{77}$Ge} and {$^{77m}$Ge} decays in enriched detectors of \textsc{Gerda} Phase\,II.
The reduction by consecutive application of active background rejection techniques is shown.
The spectra are normalized to represent the spectral contribution per decay, the binning is 20 keV.
The inset shows the R$_{90}$ distribution for events around {$Q_{\beta\beta}$}.
It is used to model the multi-site event rejection by PSD. 
}
\label{fig:decay}
\end{figure*}
%%%%%%%%%%%%%%%%%%%%%%%%%%%%%%%%%%%%%%%%%%%%%%%%%%%%%%%%%%%%%%%%%%%%%%%%%%%%%%%%

As mentioned above, the data-driven NeutronHP models based on the ENDF/B-VII data libraries are used for the tracking of neutrons below 20~MeV. 
The error due to the interpolation of adjacent data points is smaller than a few percent \cite{geant_manual}. 
It has been shown in \cite{Lemrani:2006dq} that the agreement for the neutron propagation between Geant4 and the MCNPX code \cite{mcnpx}, which uses tabulated data from the alternative JENDL database \cite{jendl}, is better than 20\%, and often within 10\%. 
Taking also into account the study of \cite{Pritychenko:2014ida}, the global uncertainty for neutron propagation due to the \textsc{Geant4} NeutronHP models is estimated here to be 20\%. 
When combining the two terms in quadrature, the global systematic uncertainty for isotope production mediated by muon induced neutrons results to be 35\%, provided that the cross sections for the relevant isotopes are included in the ENDF/B-VII library.\footnote{The uncertainty is slightly smaller than used in \cite{NIMA}, given the additional studies on muon induced neutron yields of \cite{Abe:2009aa,Reichhart:2013xkd}.}

The systematic uncertainty on the production yield is substantially larger for those nuclei that are mostly produced by other mechanisms which do not involve neutrons, as lepto-nuclear reactions, direct spallation or high-energy proton scattering. 
This can be up to a factor of four for germanium targets and for the product nuclei of interest for this work \cite{Wei:2017hkq,Cebrian:2017oft}.

\subsection{Active background rejection}
\label{subsec:vetos}

Apart from passive background reduction in terms of graded passive shielding and extensive material selection efforts, active background rejection is a key feature of the \textsc{Gerda} experimental approach.
The electrons released in {$^{76}$Ge} double beta decay will deposit their energy within short range in one single site in one HPGe detector.
Background events exhibiting a different topology can be efficiently rejected by the means of active background rejection techniques \cite{NeutrinoProceedings}.

Events with simultaneous energy depositions in multiple HPGe detectors are rejected by detector anti-coincidence (AC).
A realistic offline trigger threshold of 10~keV in the AC condition of simulated data is used.
Analogously events with coincident energy deposition in the LAr surroundings are suppressed by the Phase\,II scintillation light read-out system (LAr veto).
For the sake of computing power a simplified effective model is used.
Events are rejected if a coincident energy deposition in the LAr volume contained by the light instrumentation exceeds 150~keV.

The finite drift time of charge carriers in HPGe detectors allows to discriminate events differing from single-site bulk energy depositions by pulse shape discrimination (PSD) \cite{PSD}.
The multi-site event rejection of PSD is modeled by a simple post-simulation parametrization described in \cite{R90}. 
The parameter R$_{90}$, defined as the radius from the barycenter of energy depositions containing 90\% of the deposited energy, is calculated per HPGe detector.
Energy depositions exceeding a size of 2~mm in R$_{90}$ are classified as rejected. 
The signal acceptance for simulated {$0\nu\beta\beta$} events with energy deposition at {$Q_{\beta\beta}$} is 97.3\% (see inset in Figure~\ref{fig:decay}).
This approximative model allows to draw fast and simple conclusions on the multi-site rejection performance by PSD without the need for sophisticated pulse shape simulations.

The parameters used in this simplified modeling of the active background rejection are chosen to represent a conservative estimate in comparison to experiences with the \textsc{Gerda} setup.
Ongoing efforts for next generation of experiments aim for improved background suppression performance by e.g. increased LAr light yield, improved photon detection and low threshold/low noise HPGe detector read-out electronics.
The obtained results hence represent a conservative baseline for future experiments.

%----------------------------------------------------------------------------------
%---------------------------------------------------------------------------- Ge 77
\section{Active suppression of {$^{77(m)}$Ge} decays}
\label{sec:ge77}

With a Q-value of about 2.7 MeV both the $\beta$-decays of {$^{77}$Ge} and {$^{77m}$Ge} contribute to the background of experiments searching for $0\nu\beta\beta$ decay with {$^{76}$Ge}.
The simplified decay scheme is depicted in Figure~\ref{fig:decay}.
The half-life is 11.2~h for {$^{77}$Ge} and 53.7~s for {$^{77m}$Ge} \cite{Ge77}.
With a probability of (19$\pm$2)\% {$^{77m}$Ge} undergoes internal transition to {$^{77}$Ge} and emits a $\gamma$ with 160~keV.
The direct transition between the ground states of {$^{77}$Ge} and of its daughter nucleus {$^{77}$As} is spin-suppressed.
All {$^{77}$Ge} $\beta$-decays are followed by $\gamma$ emission from de-excitation of {$^{77}$As} excited states.
A minimum of 195 keV corresponding to the first excited state of {$^{77}$As} is released in coincidence.
This results in a high chance to reduce this background by state-of-the-art active background rejection techniques.
Vice-versa {$^{77m}$Ge} decays in (58$\pm$4)\% via pure ground-state decay without emitting additional $\gamma$'s.
A single $\beta$ released in the bulk of a HPGe detector represents a similar topology as double beta events.
Hence, a large fraction of this background is expected to be irreducible by the current background rejection techniques based on prompt coincidences.
Nevertheless, the comparably short half-life of {$^{77m}$Ge} opens up the possibility for rejection by delayed coincidence cuts as discussed in Section~\ref{sec:cuts}. 

{$^{77}$Ge} and {$^{77m}$Ge} decays have been simulated in enriched HPGe detectors of the \textsc{Gerda} Phase\,II array.
The obtained spectra are shown in Figure~\ref{fig:decay}.
The inset shows the R$_{90}$ distribution in comparison with the single-site dominated distribution for $0\nu\beta\beta$ decays. 
As expected from decay scheme considerations, the suppression by active background rejection is large for {$^{77}$Ge} only.
The shape of the final spectrum after application of AC, LAr veto and PSD multi-site rejection reveals the contributions from $\beta$'s to different levels of the {$^{77}$As} daughter nucleus.
Different $\gamma$ emissions from subsequent de-excitations are visible in the unsuppressed spectrum prior cuts.
Only minor suppression is achieved for {$^{77m}$Ge}.
It is driven by a subdominant transition via the 215~keV state of {$^{77}$As}.
The unsuppressed 160~keV peak corresponds to the internal transition, the subsequent decay is taken into account in the {$^{77}$Ge} spectrum.

The resulting spectral contributions at $Q_{\beta\beta}\pm$100~keV are summarized in Table~\ref{tab:param}.
Generally $\beta$'s are contained in a single detector.
The dominance of $\beta$'s in the {$^{77m}$Ge} energy release manifests in a larger initial contribution at {$Q_{\beta\beta}$} and less rejection power by active background suppression based on prompt coincidences.  

%%%%%%%%%%%%%%%%%%%%%%%%%%%%%%%%%%%%%%%%%%%%%%%%%%%%%%%%%%%%%%%%%%%% table %%%%%
\begin{table}
\centering
\caption{
Spectral contribution of {$^{77}$Ge} and {$^{77m}$Ge} decays at $Q_{\beta\beta}\pm$100~keV before and after application of detector anti-coincidence (AC), liquid argon (LAr) veto and PSD multi-site rejection (R$_{90}).$ 
}
\label{tab:param}
\begin{tabular*}{\columnwidth}{@{\extracolsep{\fill}}lcc@{}}
\hline 
  & {$^{77}$Ge}              &  {$^{77m}$Ge}           \\  
  &  [10$^{-5}$ keV$^{-1}$]  &  [10$^{-5}$ keV$^{-1}$] \\   
\hline 
  prior cuts                            & 12.6  & 23.3 \\
  after AC                              & 10.5  & 22.4 \\ 
  after AC and LAr veto                 &  7.5  & 21.2 \\
  after AC, LAr veto and R$_{90}$ cut   &  1.2  & 17.0 \\  
\hline 
\end{tabular*}
\end{table}
%%%%%%%%%%%%%%%%%%%%%%%%%%%%%%%%%%%%%%%%%%%%%%%%%%%%%%%%%%%%%%%%%%%%%%%%%%%%%%%%

%----------------------------------------------------------------------------------
%----------------------------------------------------------------------------- MUON

\section{Cosmic muon induced {$^{77(m)}$Ge} production}
\label{sec:ncapture}

\begin{figure}[bt]
\centering
\includegraphics[width=\columnwidth]{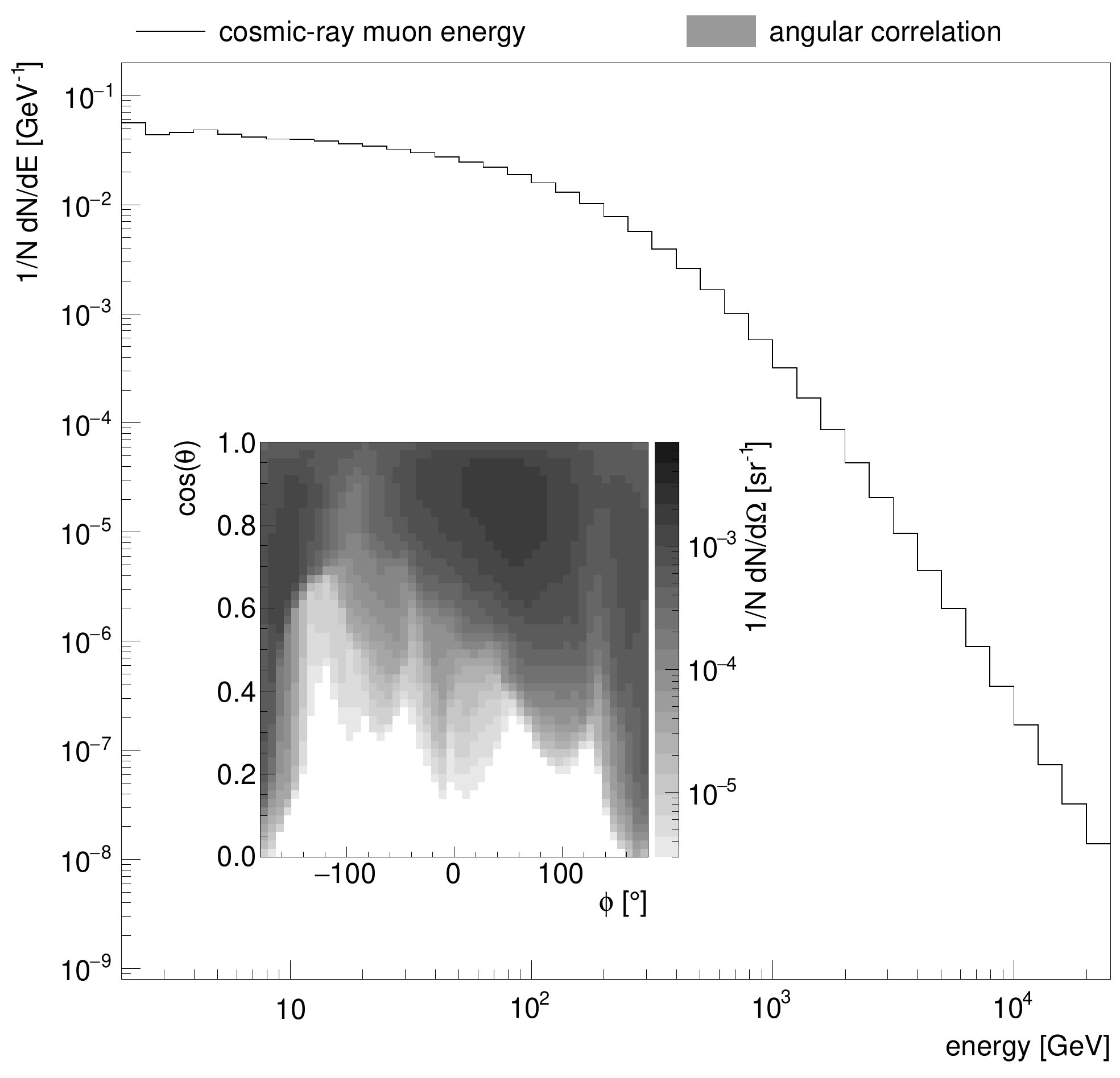}
\caption{
Energy spectrum and azimuth/zenith angle correlation of muons produced by the \textsc{Musun} \cite{Musun} code to imping the \textsc{Gerda} setup.
}
\label{fig:muons}
\end{figure}

\begin{figure}[bt]
\centering
\includegraphics[width=\columnwidth]{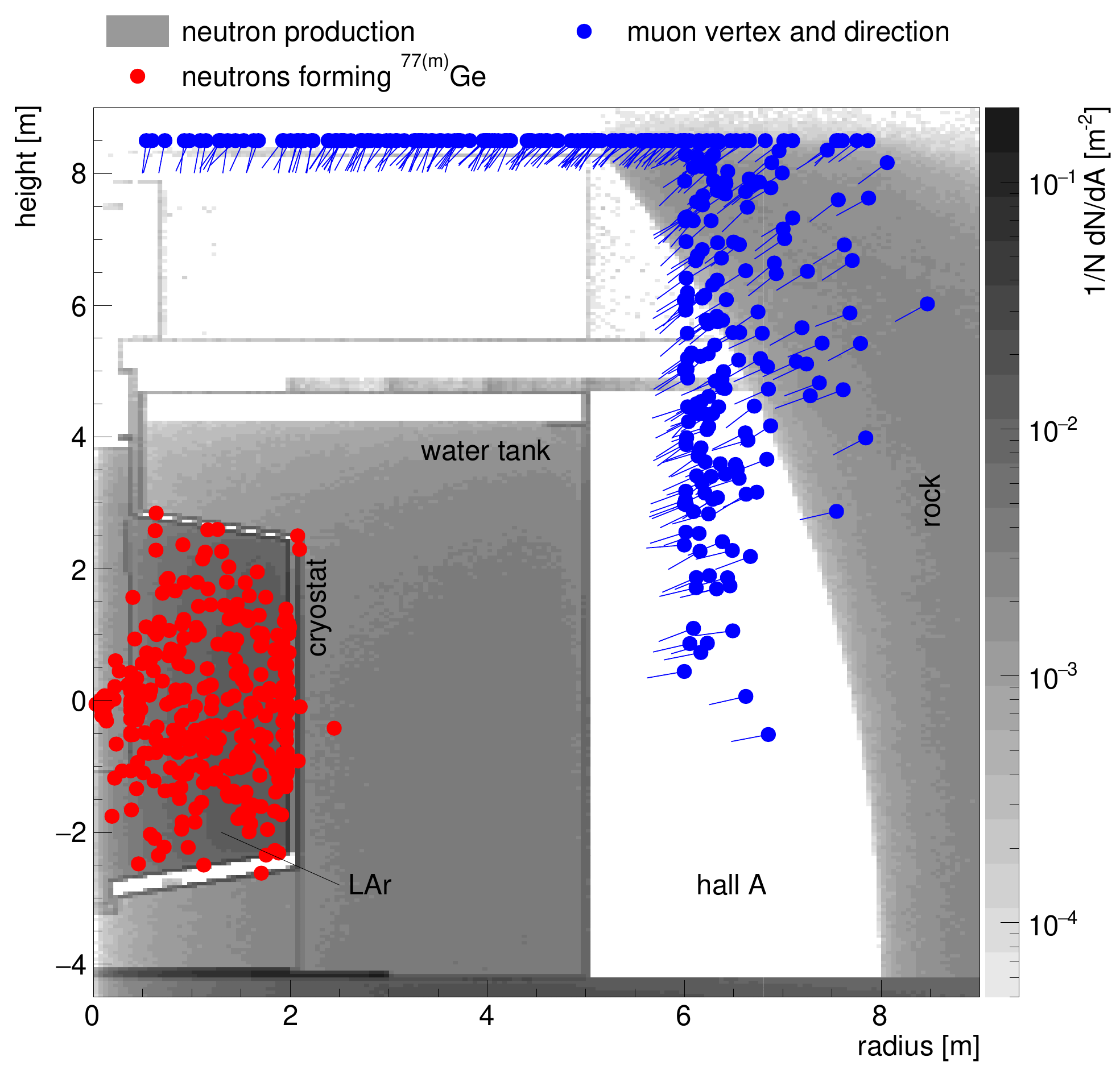}
\caption{
Cosmic muon induced neutron production in the \textsc{Gerda} geometry.
The production vertices for neutrons being captured by {$^{76}$Ge} and the corresponding muons are highlighted. 
}
\label{fig:neutrons}
\end{figure}

The cosmic muon induced production of radioactive isotopes was studied by the Monte Carlo simulation of muons impinging the \textsc{Gerda} setup.
Their distribution in energy and direction was simulated with the \textsc{Musun} \cite{Musun} Monte Carlo code and is shown in Figure~\ref{fig:muons}.
The mean energy of the muons is 270~GeV.
The azimuth and zenith angle distributions follow the profile of the Gran Sasso mountain.
The code has been validated against measurements by the LVD experiment \cite{lvd} situated next to \textsc{Gerda} in Hall A of the \textsc{Lngs} underground laboratory.
In total 1$\cdot$10$^8$ muons were released from a 12x12x13~m$^3$ sized box around the \textsc{Gerda} setup.\footnote{No sufficient range to generate hadronic showers in the surrounding Gran Sasso rock is included (see Figure~\ref{fig:neutrons}). 
Simulations including additional meters of rock led to similar results, but less statistics due to increased CPU time.}
This corresponds to a lifetime of 43.4~yr or 1544~kg$\cdot$yr of \textsc{Gerda} Phase\,II exposure.

%%%%%%%%%%%%%%%%%%%%%%%%%%%%%%%%%%%%%%%%%%%%%%%%%%%%%%%%%%%%%%%%%%%% figure %%%%
%%%%%%%%%%%%%%%%%%%%%%%%%%%%%%%%%%%%%%%%%%%%%%%%%%%%%%%%%%%%%%%%%%%%%%%%%%%%%%%%

Among other particles, neutrons are produced in various processes induced by cosmic muons.
The main production channels are photon nuclear interactions and the production of secondary neutrons from inelastic neutron interactions.
Figure~\ref{fig:neutrons} shows the spatial distribution of neutron production vertices.
The muon induced neutron flux at the position of the HPGe detector array is 1.6~m$^{-2}$h$^{-1}$. 
This is less than the 5.7~m$^{-2}$h$^{-1}$ obtained in \cite{NIMA}.
The difference can be attributed to the updated \textsc{Gerda} Phase\,II setup with respect to the original design (notably, the different shape, dimensions and material of the cryostat) and to changes in \textsc{Geant4}. 

After scattering down in energy, neutrons end up to be absorbed in materials of the \textsc{Gerda} setup.
The production vertices of neutrons eventually forming {$^{77(m)}$Ge} in the germanium detectors are highlighted in Figure~\ref{fig:neutrons}.
Mainly neutrons generated in the LAr cryostat walls or in the LAr volume contribute to the {$^{77(m)}$Ge} production.
The water tank efficiently shields the detector array from other neutrons.
All muons initiating the production of {$^{77(m)}$Ge} point towards the cryostat volume.
Proper tracking capability for muons might allow to identify muons associated with neutron production in the inner part of the experiment.

%%%%%%%%%%%%%%%%%%%%%%%%%%%%%%%%%%%%%%%%%%%%%%%%%%%%%%%%%%%%%%%%%%%% figure %%%%
%%%%%%%%%%%%%%%%%%%%%%%%%%%%%%%%%%%%%%%%%%%%%%%%%%%%%%%%%%%%%%%%%%%%%%%%%%%%%%%%

\begin{figure*}[t]
\centering
\includegraphics[width=\textwidth]{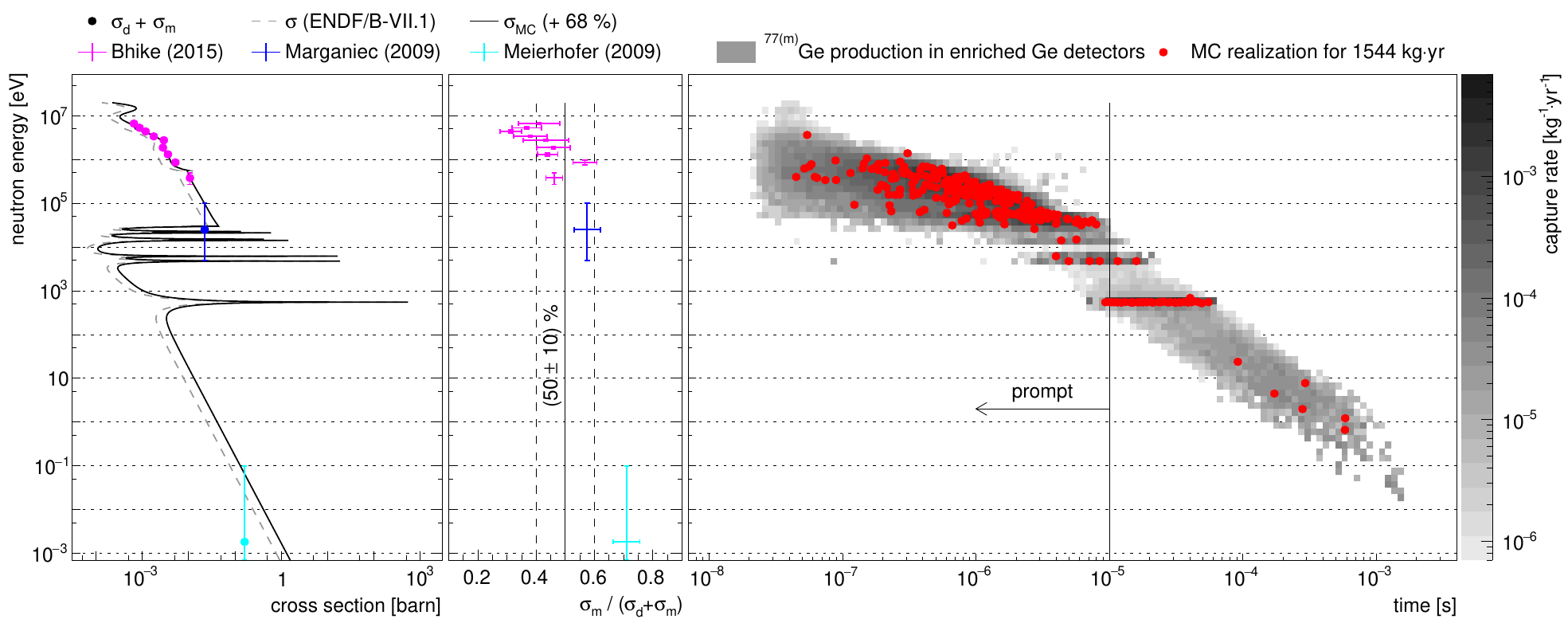}
\caption{
Left: Recent cross section data for {$^{77}$Ge} and {$^{77m}$Ge} production by neutron capture on {$^{76}$Ge}.
To account for full {$^{77m}$Ge} production the cross section used in the simulation was increased by 68\% compared to the plain ENDF-based one.
Middle: Ratio of $\sigma_m$ to total $\sigma_d$~+~$\sigma_m$ cross section. 
An estimate of (50$\pm$10)\% was used in the analysis.
The error bars are used to indicate the range of continuous neutron energies.
Right: Muon induced {$^{77(m)}$Ge} production rate over neutron energy and time. 
The Monte Carlo realization corresponds to 1544~kg$\cdot$yr of \textsc{Gerda} Phase\,II exposure.
}
\label{fig:capture}
\end{figure*}

The cross section for neutron capture on {$^{76}$Ge} has been recently measured at different energies \cite{Meierhofer, Marganiec, Bhike}.
The evaluations are based on the determination of $\gamma$ intensities involved in the {$^{77}$Ge} and {$^{77m}$Ge} decay.
The directly accessible quantities are $\sigma$ and $\sigma_m$. 
The cross section $\sigma$ includes the direct {$^{77}$Ge} production cross section $\sigma_d$ and (19$\pm$2)\% of internal transitions from {$^{77m}$Ge}, whereas $\sigma_m$ describes the {$^{77m}$Ge} production only.
The G4NDL library responsible for neutron capture on {$^{76}$Ge} in \textsc{Geant4} is based on ENDF/B-VII-1 tabulated data, which does not provide $\sigma_d$~+~$\sigma_m$ for total {$^{77(m)}$Ge} production.
The ratio of $\sigma_m$ to $\sigma_d$~+~$\sigma_m$ measured at different neutron energies is shown in Figure~\ref{fig:capture}.
With increasing neutron energy the cross sections favor {$^{77}$Ge} over {$^{77m}$Ge} production, as higher neutron energies may result in higher excited states \cite{Bhike}.
Considering the neutron energies involved in the in-situ {$^{77(m)}$Ge} production a constant ratio for $\sigma_m$ to $\sigma_d$~+~$\sigma_m$ of (50$\pm$10)\% is used in this analysis.
This implies that the plain \textsc{Geant4} cross section underestimates {$^{77(m)}$Ge} production by 68\%.
The Monte Carlo cross section $\sigma_{MC}$ was artificially increased in the performed simulation to account for this.
It is depicted in Figure~\ref{fig:capture}.
Measurements representing $\sigma_d$~+~$\sigma_m$ agree reasonably well.

%%%%%%%%%%%%%%%%%%%%%%%%%%%%%%%%%%%%%%%%%%%%%%%%%%%%%%%%%%%%%%%%%%%% figure %%%%
%%%%%%%%%%%%%%%%%%%%%%%%%%%%%%%%%%%%%%%%%%%%%%%%%%%%%%%%%%%%%%%%%%%%%%%%%%%%%%%%

Neutron interaction cross-sections of materials in the germanium array vicinity define the spatial and temporal distribution of muon induced neutrons entering the array, especially at low neutron energies where the cross section for neutron capture on $^{40}$Ar dominates.
Folding the neutron distribution with the neutron capture on {$^{76}$Ge} cross section the {$^{77(m)}$Ge} production rate was obtained. 
Figure~\ref{fig:capture} shows the resulting capture rate in neutron energy vs. time.
A large fraction of the {$^{77(m)}$Ge} production appears to be prompt (<~10~$\mu$s) after the muon hits the experiment.
Only the resonance at 550~eV has strong contribution to delayed captures.
The cross section at thermal energies plays a minor role.
The simulation yields a production rate of muon induced {$^{77(m)}$Ge} in \textsc{Gerda} of (0.21$\pm$0.01)~nuclei/(kg$\cdot$yr).
It represents the main production channel\footnote{Non-cosmogenic {$^{77m}$Ge} production by radiogenic neutrons was found to be one order of magnitude lower.} and allows verification in \textsc{Gerda} and next generation germanium experiments.\footnote{The comparably short half-life of {$^{77m}$Ge} and the presence of a metastable state of {$^{77}$As} at 475~keV with a half-life of 116~$\mu$s in the {$^{77}$Ge} decay open up possibilities to separately measure the {$^{77}$Ge} and {$^{77m}$Ge} production rate.}
In combination with the results from Section~\ref{sec:ge77} this leads to a background contribution of (1.5$\pm$0.2)$\cdot10^{-6}$ cts/(keV$\cdot$kg$\cdot$yr) from {$^{77}$Ge} and (1.8$\pm$0.4)$\cdot10^{-5}$ cts/(keV$\cdot$kg$\cdot$yr) from {$^{77m}$Ge}.\footnote{Internal transitions from {$^{77m}$Ge} to {$^{77}$Ge} are taken into account.}

%----------------------------------------------------------------------------------
%-------------------------------------------------------------------------- DELAYED
\section{{$^{77m}$Ge} rejection by delayed coincidence}
\label{sec:cuts}

%%%%%%%%%%%%%%%%%%%%%%%%%%%%%%%%%%%%%%%%%%%%%%%%%%%%%%%%%%%%%%%%%%%% figure %%%%
\begin{figure*}[t]
\centering
\includegraphics[width=\textwidth]{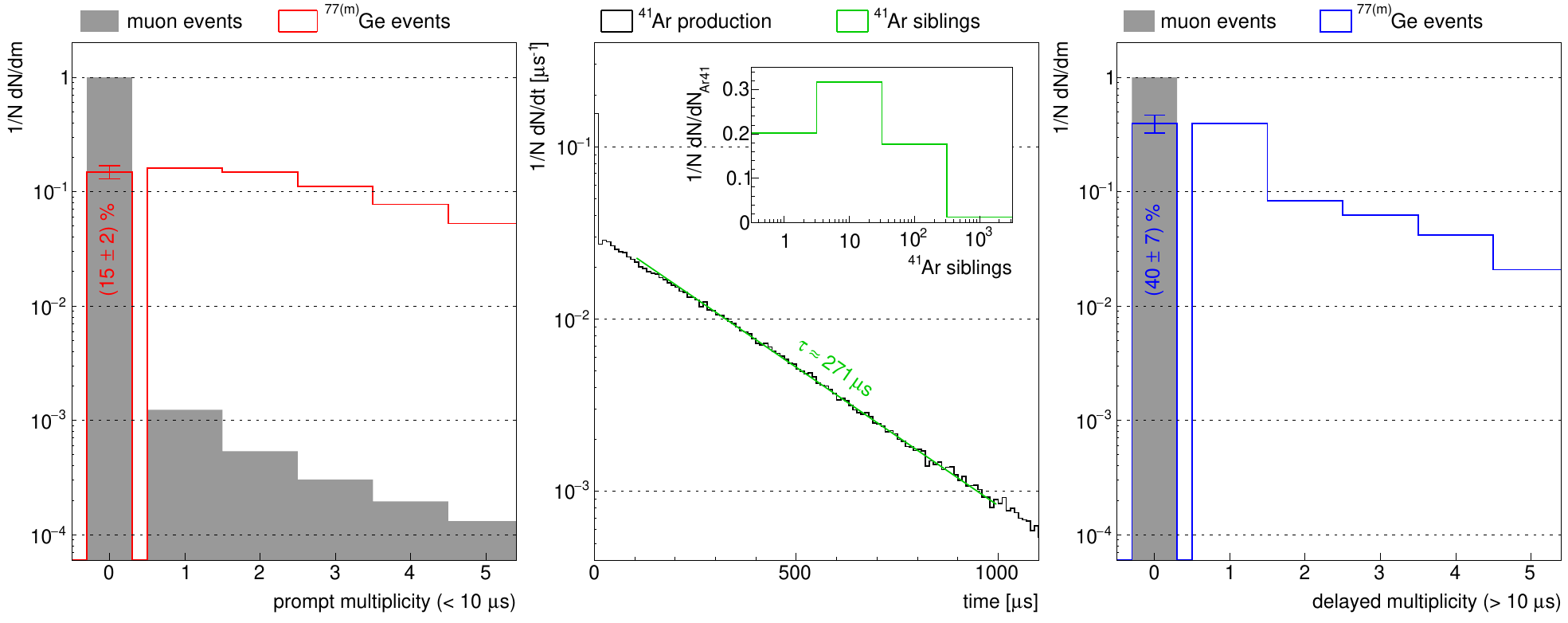}
\caption{
Left: Prompt germanium detector multiplicity for events tagged as muon event and events with {$^{77(m)}$Ge} production. 
Only (15$\pm$2)\% of the {$^{77(m)}$Ge} events do not show prompt coincidence between muon veto and germanium detectors.
Middle: Timing of {$^{41}$Ar} production. 
The inset shows the number of accompanying {$^{41}$Ar} siblings for {$^{77(m)}$Ge} events.
Right: Delayed (>~10~$\mu$s) multiplicity for events not showing prompt coincidences in the first place.
}
\label{fig:rejection}
\includegraphics[width=\textwidth]{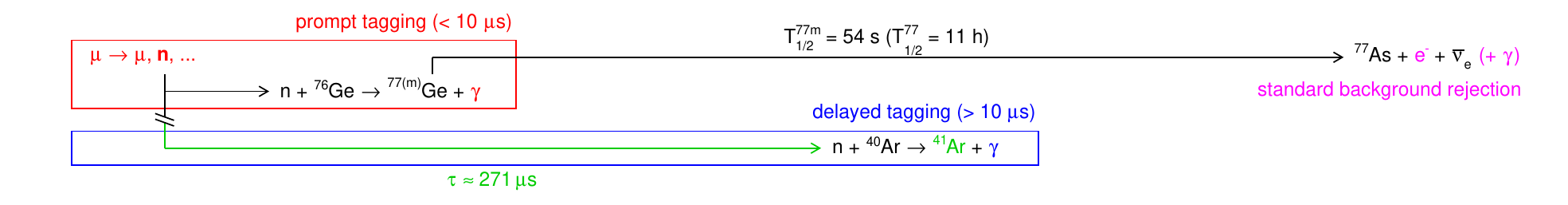}
\caption{
Sequence of the delayed coincidence cut based on tagging by prompt and delayed coincidences between muon veto and germanium detector signals.
The delayed signals are mainly provided by capture of thermalized neutrons in the surrounding LAr.
The comparably short half-life of {$^{77m}$Ge} allows for suppression by a delayed coincidence cut after receiving either a prompt or delayed tagging signal.
The accompanying $\gamma$'s in the subsequent decay are essential in the suppression by standard background rejection techniques based on prompt coincidences.
}
\label{fig:scheme}
\end{figure*}
%%%%%%%%%%%%%%%%%%%%%%%%%%%%%%%%%%%%%%%%%%%%%%%%%%%%%%%%%%%%%%%%%%%%%%%%%%%%%%%%

The strong single beta component of {$^{77m}$Ge} is irreducible with background rejection techniques based on prompt coincidences.
However, the comparably short half-life of 53.7~s opens up possibilities to reduce this background by delayed coincidence cuts.
Neutron capture leaves the daughter nucleus in a highly excited state, according to the available nuclear binding energy and level scheme.
Its de-excitation results in the release of $\gamma$'s.
In \textsc{Geant4} this is represented by tabulated $\gamma$ emission data \cite{geant_manual}. 
According to Figure~\ref{fig:capture} most captures on {$^{76}$Ge} appear shortly after the actual muon is detected by the \textsc{Gerda} muon veto system.
The read-out scheme will not allow to register separate events in muon veto and germanium detectors if within <~10~$\mu$s.
Figure~\ref{fig:rejection} shows the prompt single event multiplicity registered in the germanium detector array for muon events and for muon events with {$^{77(m)}$Ge} production.
In addition to the prompt signal from the muon induced shower itself, the $\gamma$ emission from inelastic neutron interactions in and around the array contributes to the enhanced multiplicity for the latter one.
The efficiency to tag events with {$^{77(m)}$Ge} production by prompt coincidences (multiplicity $\geq$~1) between muon veto and germanium array is (85$\pm$2)\%.
General muon events show this signature with a rate of 1$\cdot$10$^{-4}$~s$^{-1}$.
This allows to suppress the {$^{77m}$Ge} background by implementing a delayed coincidence cut that removes several {$^{77m}$Ge} half-lives of data after a prompt tagging signal.
A dead-time of 6~min %($\sim$5$\cdot \tau_{Ge77m}$) 
results in a reduction of the {$^{77m}$Ge} background to (3.0$\pm$0.7)$\cdot10^{-6}$ cts/(keV$\cdot$kg$\cdot$yr) at an acceptable life time reduction of <~4\%.\footnote{The efficiency of the \textsc{Gerda} muon veto system is (99.2$^{+0.3}_{-0.4}$)\% \cite{MuVeto}.}

The number of events featuring a prompt coincidence is larger than the prompt {$^{77(m)}$Ge} production itself.
Part of the registered signals originate from other processes triggered by the same incident muon, including accompanying neutron captures in the array surroundings.
These sibling captures can be used to improve the tagging efficiency for muon events with accompanying isotope production.
{$^{41}$Ar} constitutes the majority of these sibling isotopes.
A significant fraction of it is formed by capture of thermalized neutrons.
The timing for its production is shown in Figure~\ref{fig:rejection}.
It is predominantly described by a capture time of about 271~$\mu$s of thermalized neutrons in LAr.
The inset shows the number of {$^{41}$Ar} siblings for {$^{77(m)}$Ge} events, in some cases up to several hundreds of siblings are formed.
Their production significantly contributes to prompt and especially delayed signals seen in the germanium array.
(60$\pm$7)\% of {$^{77(m)}$Ge} events not showing the prompt signature still exhibit a germanium signal at >~10~$\mu$s.
Figure~\ref{fig:rejection} shows the single event multiplicity of these delayed events.
This delayed coincidence can be used as further signature to tag muon events with accompanying isotope production.
Again the implementation of 6~min dead-time after a delayed tagging coincidence allows to further reduce the {$^{77m}$Ge} background to (1.2$\pm$0.5)$\cdot10^{-6}$ cts/(keV$\cdot$kg$\cdot$yr) with negligible life time reduction.\footnote{For a coincidence window of about 1~ms the introduced dead-time is at ${O}(0.1)$\%.} 

The sequence of processes producing prompt and delayed coincidences between muon veto and germanium detectors is depicted in Figure~\ref{fig:scheme}.
In many cases the prompt signal in the germanium detectors is supported by the neutron capture on {$^{76}$Ge} itself, while for the delayed tagging capture of thermalized neutrons in the surrounding LAr is essential.
A systematically underestimated isotope production rate by muon induced neutrons as discussed in Section~\ref{subsec:mc} could partly compensate.
Larger production rates would manifest in an increased chance to tag {$^{77(m)}$Ge} events.
Additional reduction could be achieved by facilitating the LAr instrumentation to increase the efficiency to detect muon events with accompanying isotope production by observation of sibling captures.
Due to the much longer life-time delayed coincidence cuts will almost not affect the background by ground state {$^{77}$Ge}.
The combined {$^{77(m)}$Ge} background after cuts is (2.7$\pm$0.3)$\cdot10^{-6}$ cts/(keV$\cdot$kg$\cdot$yr).
However, any improvement in active background suppression based on prompt coincidences like PSD and LAr veto will result in a reduction of the {$^{77}$Ge} contribution to this number.

%----------------------------------------------------------------------------------
%--------------------------------------------------------------------------- OTHERS
\section{Other muon induced backgrounds}
\label{sec:others}

Other radioisotopes formed in cosmic muon interactions have only a minor contribution to the background in the search for {$0\nu\beta\beta$} decay of {$^{76}$Ge}.
Different backgrounds due to internal and external occurring subsequent decays were studied in \cite{NIMA}.
Only sources exceeding $10^{-6}$~cts/(keV$\cdot$kg$\cdot$yr) were reported.
Simple decay scheme considerations allow to draw qualitative predictions concerning a further suppression of these backgrounds by the \textsc{Gerda} active background rejection strategy and the proposed delayed coincidence cuts.
Internal decays of the Ga isotopes {$^{74}$Ga} and {$^{76}$Ga} will face a strong suppression from coincident $\gamma$'s of more than 500~keV released in their $\beta$-decays \cite{Ga74, Ga76}.\footnote{The emission of $\gamma$'s with 2040.7~keV from {$^{76}$Ga} $\beta$-decay mentioned in \cite{Tornow} will face strong suppression by multiple means.
The internal $\beta$-decay as well as $\gamma$'s released in the cascade are able to provide prompt coincidences.}
Due to its non-negligible ground state $\beta$ transition probability {$^{75}$Ga} will exhibit only marginal suppression.
Nevertheless, the short half-lifes of both {$^{75}$Ga} of 126~s and {$^{76}$Ga} of 33~s allow reduction by delayed coincidence.
Background originating from  {$^{68}$Ge}/{$^{68}$Ga} and {$^{69}$Ge} electron capture/$\beta^{+}$ decays will be suppressed by coincidences in the surrounding LAr or other HPGe detectors \cite{GeGa68, Ge69}.
$\beta$-decays of Ar and Cl isotopes in the surrounding LAr are negligible compared to contributions by naturally present background from e.g.  {$^{42}$K} decays \cite{minishroud}.
The high energy $\gamma$'s from  {$^{38}$Cl} and  {$^{40}$Cl} decays will be largely suppressed by the coincident energy deposit of the initial $\beta$ in the LAr \cite{Cl38, Cl40}.

The prompt muon induced background for the \textsc{Gerda} experiment has been studied in \cite{NIMA} and \cite{MuVeto}.
In addition to the muon veto, the active background rejection by AC, LAr veto and PSD further reduces this background.

%----------------------------------------------------------------------------------
%---------------------------------------------------------------------- CONCLUSIONS
\section{Conclusions and prospects for future experiments}
\label{sec:concl}

Distinct features in the production and the decay of radioactive isotopes from cosmic muon interactions allow for a strong reduction of these backgrounds in experiments exploiting an active background rejection strategy.
The initial {$^{77(m)}$Ge} background is (4.0$\pm$0.4)$\cdot10^{-5}$ cts/(keV$\cdot$kg$\cdot$yr).
State-of-the-art active background rejection techniques in combination with a simple delayed coincidence cut based on tagging muon events with accompanying isotope production by prompt and delayed coincidences between muon veto and germanium array will allow a reduction by more than one order of magnitude to (2.7$\pm$0.3)$\cdot10^{-6}$ cts/(keV$\cdot$kg$\cdot$yr) at an acceptable life-time loss of <~4\%.
The systematic uncertainty on this number is 35~\% related to the muon induced isotope production in \textsc{Geant4}.
The reduction obtained by the proposed strategy is based on conservative parameters for germanium detector and LAr scintillation light read-out.
It amounts to a reduction in muon flux of more than one order of magnitude, which is usually provided by deeper underground laboratories.
Taking the equivalent vertical depth definition of \cite{MeiHime} and attributing the reduction linearly to a decrease in muon flux, this corresponds to a Gran Sasso depth exceeding 5000~m.w.e. for {$^{77(m)}$Ge}. 

In a first stage the next generation experiment \textsc{Legend} will operate about 200~kg of enriched germanium detectors in the \textsc{Gerda} infrastructure at \textsc{Lngs} \cite{LegendMedex}.
Similar or better background rejection capabilities will result in a background contribution comparable to the one described in this paper.
This is well below the background goal of the \textsc{Legend}-200 phase at about $10^{-4}$~cts/(keV$\cdot$kg$\cdot$yr).
With a design exposure of about 1~ton$\cdot$yr for \textsc{Legend}-200 will give the unique opportunity to determine the muon induced {$^{77(m)}$Ge} production in-situ at \textsc{Lngs} and thereby verify the predicted rate of 0.21~nuclei/(kg$\cdot$yr).

For the following \textsc{Legend}-1k phase a new facility holding up to 1~ton of HPGe detectors is conceived.
Already the {$^{77(m)}$Ge} background obtained for the \textsc{Gerda} setup, exploiting the current active background rejection techniques and a simple delayed coincidence cut, is below the aspired background index goal of about $10^{-5}$~cts/(keV$\cdot$kg$\cdot$yr).
Further reduction can be achieved provided that rigorous measures are taken during the design stage by geometry optimization and material selection with respect to neutron production and propagation.

The possibilities to actively reduce the muon induced background described in this paper equals a virtual increase in \textsc{Lngs} depth and opens the way for next generation experiments with ultra low background requirements.

\begin{acknowledgements}

The authors would like to thank all members of the \textsc{Gerda} collaboration, and especially R.~Brugnera, P.~Grabmayr, Y.~Kerma\"idic, A.~Lazzaro and A.J.~Zsigmond for their helpful comments and the prolific discussions.
We gratefully acknowledge the support by the Collaborative Research Center "Neutrinos and Dark Matter in Astro- and Particle Physics'' (SFB 1258).

\end{acknowledgements}

%\appendix
%\section{Appendix section}\label{app}
%\subsection*{Appendix subsection}

%--------------------------------------------------------------- references ----
\bibliographystyle{plain}
\bibliography{references}

%\begin{thebibliography}{9}
%\bibitem{r1}
%S. Chekanov et al. (ZEUS Collaboration), Eur. Phys. J. C \textbf{42}, 1 (2005)
%\end{thebibliography}

\end{document}